\documentclass[a4paper,twoside]{article}
\newcommand{\affil}[1]{$^{\rm #1}$}
\usepackage{graphicx}
\date{} 
\title{\large\bf\flushleft 
Citations to Australian Astronomy: 5 and 10 Year Benchmarks}
\author{\parbox{\textwidth}{\flushleft
\vspace{-0.5cm}
{\it 
Katherine H.\ Kenyon\affil{A,B}, 
Arjun Paramasivam\affil{A,C}, 
Jiachin Tu\affil{A,C}, 
Albert Zhang\affil{A,D}, and
Alister W. Graham\affil{A}}\\
\vspace{0.4cm}
{\small 
\affil{A}\,Centre for Astrophysics and Supercomputing, Swinburne 
University of Technology, Hawthorn, Victoria 3122, Australia.\\
\affil{B}\,Canterbury Girls' Secondary College, Mangarra Road, Canterbury,
Victoria 3126, Australia\\
\affil{C}\,John Monash Science School, 1 Wellington Rd, Clayton Victoria 3168,
Australia.\\
\affil{D}\,Ballarat Queen's and Anglican Grammar School, 201 Forest Street,
Wendouree, Victoria 3355, Australia.}}}
\begin{document}
\twocolumn[
\begin{changemargin}{.8cm}{.5cm}
\begin{minipage}{.9\textwidth}
\vspace{-1cm}
\maketitle
%
%
\small{\bf Abstract:}

Expanding upon Pimbblet's informative 2011 analysis of career $h$-indices for
members of the Astronomical Society of Australia, we provide additional
citation metrics which are geared to a) quantifying the current performance of
b) all professional astronomers in Australia.  We have trawled the staff
web-pages of Australian Universities, Observatories and Research Organisations
hosting professional astronomers, and identified 384 PhD-qualified,
research-active, astronomers in the nation --- 132 of these are not members of
the Astronomical Society of Australia.  Using the SAO/NASA Astrophysics Data
System, we provide the three following common metrics based on publications in
the first decade of the 21$^{st}$ century (2001--2010): $h$-index,
author-normalised citation count and lead-author citation count.  We
additionally present a somewhat more inclusive analysis, applicable for many
early-career researchers, that is based on
publications from 2006--2010. 
%
Histograms and percentiles, plus top-performer lists, are presented for each 
category. 
Finally, building on Hirsch's empirical equation, we find that the (10-year) $h$-index
and (10 year) total citation count $T$ can be approximated by the relation $h = 
(0.5+\sqrt{T})/\sqrt{5}$ for $h > \sim 5$.


\medskip
\medskip
\end{minipage}
\end{changemargin}
]
\small

\section{Introduction}

Quantitative metrics, which we all experienced through school and university
—-- and which many of us now apply to individuals in our own classes --— are common
in science.  Assessment systems based upon numerical indicators
for scientific research 
invariably include some form of bibliometric indicator of quality.  While sheer
numbers of papers were once favoured, there is thankfully now an increased
emphasis on quality rather than quantity.  `Citations' are liked (by most) for
their objective and quantitative ability to grade our most common research
product, namely our papers.  Citations reflect the global perception of the
relevance and usefulness of a paper. Such community-weighted global opinion
offers a means to avoid potentially discrepant personal opinions, which can at
times be misplaced or outdated.

The most frequently used criteria for quantifying the research impact and
visibility of astronomers are, partly for the reasons given above,
citations. While acknowledging that quality and visibility are not always
synonymous, objective measures are generally more equitable and preferable
than subjective commentary in science.  Citation-based rankings actually apply
to many aspects of our profession, including not only research papers
(Burstein 2000; Pearce 2004), but also telescope and observatory performance
(Abt 1985, 2003; Peterson 1987; Trimble 1995; Grothkopf et al.\ 2007), 
Australian Research Centres
(through the Federal Government's research quality assurance audit known as
`Excellence in Research for Australia'
(ERA)\footnote{http://www.arc.gov.au/era}), Universities (through the
`Academic Ranking of World Universities' (ARWU)\footnote{http://www.arwu.org}
conducted by the Shanghai Jiaotong University), and even countries (S\'anchez
\& Benn 2004). Not surprisingly, journals themselves, including this one, are
evaluated upon citations, with Thompson Scientific generating annual 
(Garfield 1972a,b) Impact 
Factors\footnote{http://thomsonreuters.com/products\_services/science/free/essays}.


Here we provide a transparent, Australia-wide impact assessment of astronomical
research over the past 5 and 10 years through the use of several
clearly-defined, objective citation metrics.
While a decade is an acceptable time frame for measuring stable, long-term,
levels of performance by established researchers, the 5-year interval is
additionally applicable to early-career researchers
and it better matches the time frame used by the Australian Research Council when
reviewing one's immediate past performance.  
The objective statistics and benchmarks herein are expected to be of interest
to Australian astronomers and, as also noted by Pimbblet (2011), have some
relevance to nationally competitive grant schemes.  In time, such surveys may
also acquire some historical value, in particular because they can be used to
track changes and show trends.
Moreover, this paper is in the same vein as the Federal Government's November 2011
Focusing Australia's Publicly Funded Research review which called for ``a
rigorous, transparent, system-wide Australian research impact assessment
mechanism''.

Pimbblet (2011) used the Hirsch (2005) $h$-index, sometimes mis-referred to 
as the Hirsch-index, to provide a histogram and 
``top ten'' table of individual career $h$-indices for members of the 
Astronomical Society of Australia. 
Here we also provide a community histogram and ``top ten'' table of
$h$-indices, but for $h$-indices acquired over the same time interval,
specifically, the first decade of this century.  Due to shared rankings, this
table is found to contain some $\sim$20 names.  As recognised by others, such
a metric, along with the ``total citation count'', is at some level reflective of
the performance of the teams, and facilities, that one has belonged to rather
than purely an individuals performance.  We therefore also provide
histograms, percentiles, and top-20 lists for author-normalised citations and lead-author
citations from 2001--2010 and from 2006-2010.
We have allowed one year (2011) for citations to accrue; for
comparison, Pearce (2004) allowed 6 months. 

We additionally report on the demographics of our nation's research-active,
PhD-qualified astronomers, which we have discovered now tallies nearly 400.
Contributing to this much higher than anticipated number has been the
establishment and growth of new astronomy Centres over the last decade.  We 
note that our survey includes over 100 astronomers
missing from the analysis by Pimbblet (2011) because they are not voluntary
members of the Astronomical Society of Australia.  Our survey thus represents
the largest current census of professional astronomers in the nation.

\subsection{Important caveats}

Before proceeding, it may be of interest for some to learn that within the
journal Scientometrics and the Journal of Informetrics, there is an increasing
amount of publications re-assessing the merits of the $h$-index and discussing 
{\it many} alternatives, such as the $g$-index (Egghe 2006) or the $A, R$ and $AR$ 
indices (Jin et al.\ 2007), which can also have their cons.
A prominent critical review of the $h$-index can be found in a 
report by the joint Committee on Quantitative Assessment of Research (Adler
et al.\ 2008), while the extensive article by 
Panaretos \& Malesios (2009) also critically reviews the
$h$-index, and alternative single metrics (see also Moed \& van Leeuwen 1996 
and Bornmann \& Daniel 2007). 
Issues raised, some by Hisrch (2005) himself, 
include the fact that references can be both favourable and not, 
the index does not discriminate between single authorship and co-authorship, 
self-citations increase with the number of publications, popular 
references may be used to flesh out Introductions even though they 
have no real connection with the essence of the paper, and something referred
to as ``cronyism'' (Meho 2007) in which those having many co-operating
scientists may receive lots of citations. 
Most recently, Balaban (2012) have further reviewed the limitations of the
$h$-index, but note that despite these the $h$-index has gained widespread
acceptance due to its simplicity.  In spite of its inadequacies, it would be
remiss of us if we did not include the $h$-index, especially given that we are
building on the study by Pimbblet (2011). 

Astronomers spend varying amounts of their office (and home) life undertaking
research, teaching, grading, instrumentation, service and administration, public outreach
and other work-related duties.  Citations are a recognition of only one of
these activities, which is not to belittle any of the other essential tasks.
Indeed, credit for performance in the other areas exist: Carrick Awards
recognise Australian university teaching, as do several Australian Museum
Eureka Prizes, which also reward public outreach and many other worthy
activities.
There is additionally the American Astronomical Society's 
Joseph Weber Award for Astronomical Instrumentation, 
the (British) Royal Astronomical Society's Jackson Gwilt medal, 
Australia Day honours for service, and others. 

Citations are in essence a reflection of relevance to the research interests
of others.  The most brilliant article in the world will not attract citations
unless there are other scientists who are (eventually) interested.  Similarly,
there may well be geniuses working in a discipline which has only a small
community of active researchers.  Their citations will not, at the present
time at least, climb to the heights of those working on say cosmology or
exoplanets (see the discussion by Abt 2006, and/or the treatise by Wouters
1999).
Citations are therefore not a perfect measure of capability but a reflection
of productivity which others are interested in.
Furthermore, although some articles are destined to never receive vast
citations, they can still contain elements of interest to some in the 
community and be worthy of publication, such as, hopefully, this one.

\section{The Astronomers} 
 
To complete the tasks mentioned above, 
we searched for astronomers currently based in Australia 
who had during the first decade of this century 
published citable articles as catalogued within the Astrophysics Data
System\footnote{http://adsabs.harvard.edu} (ADS) 
operated by the Smithsonian Astrophysical Observatory (SAO) under a 
National Aeronautics and Space Administration (NASA) grant. 
Our nation's 150+ PhD students in astronomy were however excluded to avoid
considerably skewing the following histograms and percentiles by scholars who
have not yet had 5 or 10 years of publication history.
While the choice of database can of course influence the results, the 
ADS is the preferred database of astronomers, invariably yielding more
complete results (for astronomers) 
than those obtained from subscription-based databases such as 
Scopus\footnote{http://www.scopus.com/scopus} and the Web of
Knowledge\footnote{http://www.isiwebofknowledge.com}, and free search engines
such as Google Scholar\footnote{http://scholar.google.com}.  Being a free
service, it also has the advantage that it can be checked and used by all.

Our search for Australian-based astronomers 
was conducted using the staff web-pages within: 
our nation's many Universities (we found astronomers in 21 of these, sometimes 
located in multiple departments or schools); 
the Commonwealth Scientific and Industrial Research Organisation's
(CSIRO's) Astronomy and Space Science (CASS) and Materials Science and 
Engineering (MSE) divisions; 
the Australian Defence Force Academy (through the University of New South Wales); 
the Australian Astronomical Observatory (AAO), the Perth Observatory and 
a couple of Planetariums. 
Apologetically, we did not, however, manage to include those at the Space
Weather Branch of the Bureau of Meteorology, in particular the IPS Radio and
Space Services\footnote{http://www.ips.gov.au}. 
We then used the membership list of the Astronomical Society of
Australia\footnote{http://asa.astronomy.org.au} (ASA) to 
search for missed astronomers who are {\it Full Members}, {\it Associate
Members} or {\it Fellows} of the ASA that are based in Australia and have
published astronomy articles this past decade.

The data acquired from the ADS was obtained from 
2011 December 6--9.  For accuracy, this was re-checked from 
2011 December 12--16.
Although the ADS is fairly complete, no claim to 100\% accuracy is
made.\footnote{http://doc.adsabs.harvard.edu/abs\_doc/faq.html\#complete}

Some care was taken to identify the relevant publications when a common
author's name, such as Peterson, was encountered. 
However, as noted by Pimbblet (2011), some 
errors will inevitably creep in.  
While the use of a middle initial greatly facilitated 
the exclusion of publications from extraneous individuals with the same
surname and first initial, roughly 10\% of the entries are biased high by 
a few percent due to erroneous 
citation accreditation from other individuals.  Only nine 
individuals with common names (plus a common first initial), 
such as Smith and Jones, were excluded due 
to the time that would have been required to acquire their true citation
record.  Given the comings and goings of postdoctoral researchers and 
astronomers each year in Australia, this is considered an acceptable tolerance. 
Finally, we note that some care, albeit on a best-effort basis, 
was also given to the use of multiple variants of names, 
for example Dick=Richard, Betty=Elizabeth=Liz, Bob=Robert, etc.
Unfortunately no allowances for disturbances such as long-term illnesses,
maternity/paternity leave, job/life relocation, etc.\ could be made.
Our final citation catalogue is comprised of 375 (=384-9) research astronomers
in Australia.  

\begin{figure}[ht]
\begin{center}
\includegraphics[scale=0.5, angle=0]{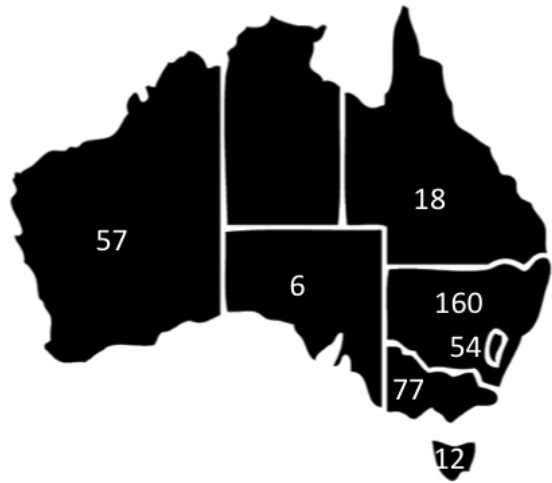}
\caption{Distribution of active, PhD-qualified, 
Australian research astronomers (2011 December).
} 
\label{Fig1}
\end{center}
\end{figure}

\begin{table*}[ht]
\caption{Distribution of Astronomers.\label{Tab1}}
\begin{center}
\begin{tabular}{lcc}
\hline
Location    & \# of Astronomers  &  \# of ASA members \\ 
\hline
CSIRO (CASS)                                 & 58   &	41 (71\%) \\
CSIRO (MSE)                                  &  1   &	 0 (0\%)  \\
The Australian National University (MSSSO)       & 40   &	27 (68\%) \\
The Australian National University (non-MSSSO)   &  9   & 	 6 (67\%) \\
The University of Sydney                     & 39   &	28 (72\%) \\
Swinburne University of Technology           & 31   &	22 (71\%) \\
The University of Western Australia (ICRAR)      & 24   &	14 (58\%) \\
The University of Western Australia (non-ICRAR)  &  5   &	 5 (100\%) \\
Australian Astronomical Observatory          & 26   &	19 (73\%) \\
Curtin University of Technology              & 25   &	14 (56\%) \\
Monash University                            & 23   &	14 (61\%) \\
Macqaurie University                          & 17   &	10 (59\%) \\
The University of New South Wales            & 13   &	11 (85\%) \\
University of Tasmania                       & 12   &	 7 (58\%) \\
The University of Melbourne                  & 10   &	 8 (80\%) \\
La Trobe University                          &  9   &	 0 (0\%)  \\
James Cook University                        &  7   &	 5 (71\%) \\
The University of Queensland                 &  7   &	 5 (71\%) \\
Australian Defence Force Academy             &  5   &	 3 (60\%) \\
The University of Newcastle                  &  5   &	 0 (0\%)  \\
The University of Adelaide                   &  4   &	 4 (100\%) \\
Perth Observatory                            &  3   &	 3 (100\%) \\
University of Southern Queensland            &  3   &	 1 (33\%) \\
University of South Australia                &  2   &	 0 (0\%)  \\
Queensland University of Technology          &  1   &	 1 (100\%) \\
University of Western Sydney                 &  1   &	 1 (100\%) \\
University of Wollongong                     &  1   &	 0 (0\%) \\
Other                                        &  3   &	 3 (100\%) \\
{\bf Total}                                  & 384  &  252 (66\%) \\
\hline
\end{tabular}
\end{center}
\end{table*}

Table~\ref{Tab1} and Figure~\ref{Fig1} provide a breakdown of where the 384
astronomers can be found. In total, 132 of these astronomers are not members
of the Astronomical Society of Australia\footnote{For those who may be
curious, and for comparison, roughly 60 of the 384 Australian astronomers are
members of the American Astronomical Society.}  according to the online
list\footnote{http://physics.usyd.edu.au/$\sim$asamail/asa\_membership/members\_html.php}.
What this means is that relative to this membership number of 252 (=384-132)
non-retired Australian-based astronomers, an additional 50\% of our nation's professional
astronomers are not members of the Astronomical Society of Australia. 
Of immediate surprise is this high number of PhD-qualified astronomers
now working in Australia.  This is in part due to a number of notable 
initiations over the last decade, including the growth of 
the Department of Physics and Astronomy at Macquarie University, 
Swinburne University of
Technology's Centre for Astrophysics --- due to a University push to increase
its research profile --- and Curtin University's Institute of Radio Astronomy
(CIRA) coupled with The University of Western Australia's International Centre
for Radio Astronomy Research (ICRAR), both established in anticipation of
the AUD\$2bn Square Kilometre Array\footnote{http://www.skatelescope.org} 
(SKA) radio telescope being built in 
Western Australia and operational from 2020.  
Collectively these three Centres employ some 80
astronomers and are training a growing number of PhD students.  Furthermore,
the Federal Government's recent Super Science
Fellowships\footnote{http://www.arc.gov.au/ncgp/ssf/ssf\_default.htm} has enabled
the recruitment of some 30 astronomy postdoctoral researchers over the last
couple of years. 
%
Given the global financial crisis over the past 3 years, which has seen a
hiring freeze at many US Universities, Australia's expansion has indeed been
fortunate.

\section{The Citation Data}

\begin{figure*}[ht]
\begin{center}
\includegraphics[scale=0.68, angle=270]{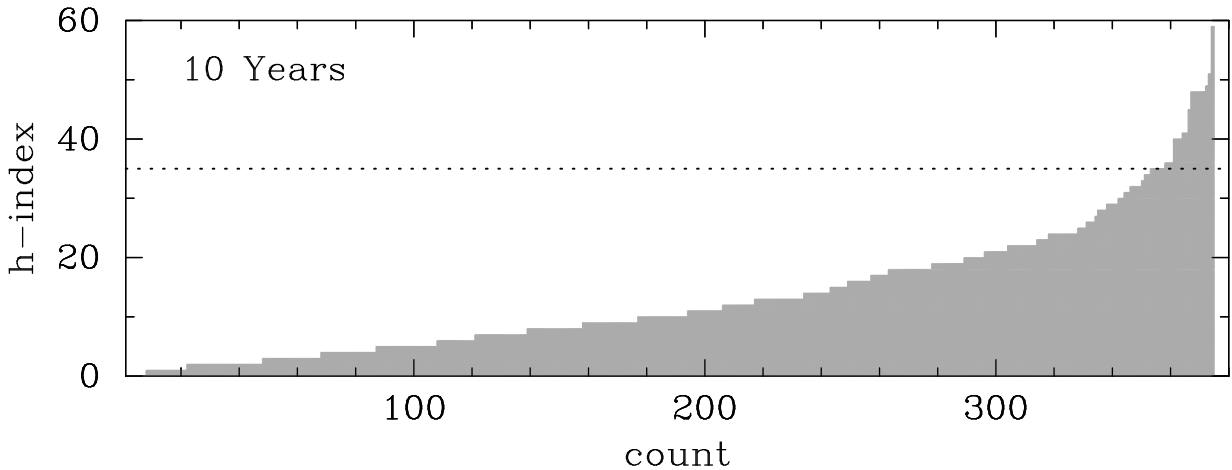}
\caption{Histogram of $h$-indices from publications in 2001--2010, 
after allowing one year (2011) for citations to accrue. 
The horizontal axis shows a running count of the ranked results. 
The dotted line at $h=35$ delineates the top 20 names in Table~\ref{Tab3}, 
and roughly corresponds to the top 5\% of the survey. 
} 
\label{Fig2}
\end{center}
\end{figure*}

Figure~\ref{Fig2} is a histogram of individual 
Hirsch (2005) $h$-indices from publications over the first decade of this 
century, i.e.\ years 2001--2010.  
Hirsch (2005) argued that this index, representing an author's 
number of papers having citation number $\ge$ $h$, 
is preferable to the total number of citations. 
In contrasting the upper echelon of career $h$-indices between 
members of the Astronomical Society of Australia and the American 
Astronomical Society (Conti et al.\ 2011), 
Pimbblet (2011) did however note that ``membership 
of very large observational programmes can boost a researcher’s h-index above
mean values''.  
Indeed, multiple author papers are known to attract more 
citations than single author articles (Abt 1984), 
possibly due to a greater advertisement of the work (Rao
\& Vahia 1986) or because of the greater input and/or
grander issues tackled.
Among the top dozen Australian names (see the Appendix, 
Table~\ref{Tab3}), 
eight are former members of the highly successful 2 degree Field (2dF) Galaxy Redshift
Survey\footnote{http://www2.aao.gov.au/2dFGRS} (e.g.\ Colless et al.\ 2001),
with one additionally belonging to the Sloan Digital Sky
Survey\footnote{http://www.sdss.org} (SDSS; e.g.\ Adelman-McCarthy et al.\
2008). Large collaborations at other wavelengths, such as 
the High Energy Stereoscopic
System\footnote{http://www.mpi-hd.mpg.de/hfm/HESS} (HESS; e.g.\ Egberts et
al.\ 2008) and the 
H${\rm I}$ Parkes All Sky Survey\footnote{http://hipass.anu.edu.au} 
(HIPASS; e.g.\ Barnes et al.\ 2001), 
can also bolster one's $h$-index. 
Furthermore, one can expect a similar positive outcome from membership of current
large observing programmes on the Australian Astronomical
Telescope\footnote{http://www.aao.gov.au} (AAT), such as WiggleZ\footnote{http://wigglez.swin.edu.au} (e.g.\
Drinkwater et al.\ 2006; Blake et al.\ 2009) and the Galaxy and Mass Assembly
(GAMA) survey\footnote{http://www.gama-survey.org} 
(e.g.\ Driver et al.\ 2008; Robotham et al.\ 2010).


\begin{figure*}[ht]
\begin{center}
\includegraphics[scale=0.65, angle=-90]{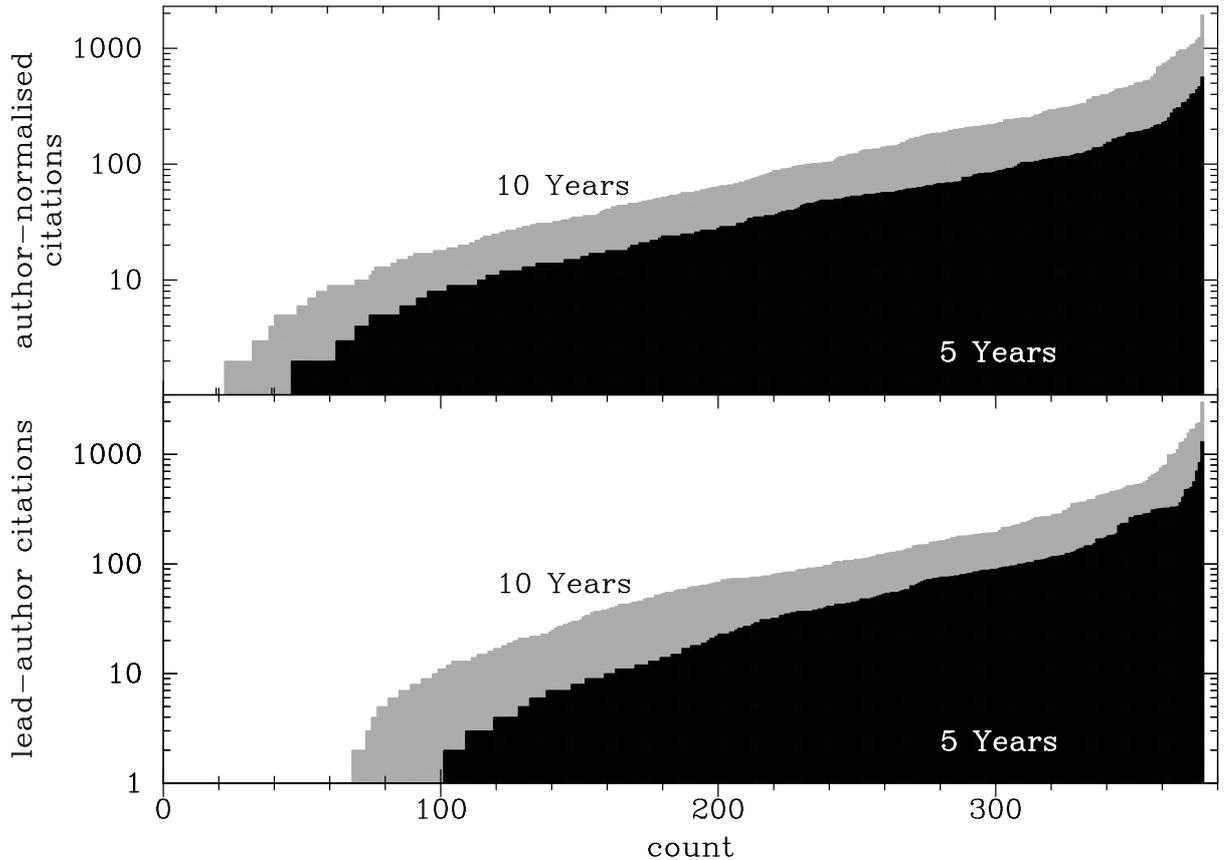}
\caption{
Top panel: Histogram of author-normalised citations 
to publications from 2001--2010 (grey) and from 2006--2010 (black). 
Lower panel: Histogram of lead-author citations
to publications from 2001--2010 (grey) and from 2006--2010 (black). 
One year (2011) was allowed to elapse for citations to accrue.
The running index along the horizontal axis reflects 
the number of astronomers in the survey. 
}
\label{Fig3}
\end{center}
\end{figure*}

We have therefore presented two additional citation metrics which reflect the fact
that authors of multiple-author papers (e.g.\ Bains et al.\ 2009; Robertson et
al.\ 2010; Waite et al.\ 2011) would have all contributed to those papers, while
the lead-author will have likely performed the bulk of the work.
Figure~\ref{Fig3} 
shows the normalized citation count (a sorting option within ADS in which each
article's total citation count is divided by the number of contributing
authors), and an individuals total citation count when they are the
lead-author (achieved by placing the carat symbol, \textasciicircum, in front
of an author's name within the ADS web-form).
The large range in citations, exemplified by the factors of 20 to 40 increase 
from the median to the highest values, and the factors of 2 to 4 increase from
the 20$^{th}$ to the top ranked individuals, 
necessitated the use of a logarithmic
scale in Figure~\ref{Fig3}.  For the curious, 
the 20 top-ranked individuals from each of the above two
citation metrics are listed in the Appendix (Tables~\ref{Tab4} to \ref{Tab5}, respectively).

The above analysis was repeated on publications from the 5 year interval
2006--2010, with the exception of the $h$-index which may suffer from `small
number' statistics.  The results are provided in Figure~\ref{Fig3}, with the 
top-ranking names again shown in the Appendix (Tables~\ref{Tab6} to \ref{Tab7}). 
Looking for anomalies, it is of interest to note, and credit, that the
stratospheric result at the top of Table~\ref{Tab7} is due to 1100+ citations to the
highly successful paper by Croton et al.\ (2006).
Many works by Asplund, such as Asplund et al.\ (2005, 2009), 
on the chemical composition of our Sun have also been incredibly useful 
and popular. 
It is also noteworthy that in the upper panel of Figure~\ref{Fig3}, 
the author-normalised citation count is seen to be remarkably log-linear from
a count above 10 (8) for the 10 (and 5) year intervals, until the final 5-10\% 
of the population is reached. 

From the histograms in Figures~\ref{Fig2} and \ref{Fig3}, 
one can readily determine the 25\%, 50\%, 75\%, 90\% and
95\% percentiles from each distribution, and these are shown in 
Table~\ref{Tab2}.
%


\begin{table}[ht]
\caption{Percentiles\label{Tab2}}
\begin{center}
\begin{tabular}{lrrrrr}
\hline
Metric  &  25\%  &  50\%  &  75\%  &  90\%  &  95\%  \\ 
\hline
$h$-index, {\it 10 yrs}    &  5   &  10   &  19    &   28  &   35  \\
normalised, {\it 10 yrs}   &  17  &  57   &  186   &  387  &  563  \\
lead-author, {\it 10 yrs}  &  10  &  59   &  164   &  424  &  650  \\
normalised, {\it 5 yrs}    &  7   &  25   &  68    &  139  &  206  \\
lead-author, {\it 5 yrs}   &  1   &  17   &  76    &  170  &  311  \\
\hline
\end{tabular} 
\end{center}
Hirsch $h$-index (2001--2010), author-normalised citation count (2001--2010
and 2001--2010) and lead-author citation count (2001--2010
and 2001--2010), after allowing one year (2011) for citations to accrue. 
Those in the 95$^{th}$ percentile, i.e.\ the top 19, 
are listed in Tables~\ref{Tab3} to \ref{Tab7}.
\end{table}


\section{Discussion}

Pimbblet (2011) recognised three issues with the histograms of career
$h$-indices that he presented: 
(i) there needs to be a calibration relative to the number
of years in the business  
--- which is why Hirsch (2005) introduced the
$m$-index\footnote{The $m$-index is essentially the $h$-index divided by the
number of years (minus 1) that one has been publishing.};  
(ii) that membership of large research programmes can inflate one's $h$-index;
and (iii) an unknown fraction of our nation's astronomers were excluded from
his analysis.  Indeed, his concluding remarks spoke to this, along with the 
instructive nature that such surveys can have for future comparisons. 

In an effort to address point (i), Pimbblet first published a table of percentiles 
for the $h$-index distribution based on one's membership class within the
Astronomical Society of Australia: a rough, and admittedly limited, measure of
seniority and thus number of years publishing.  This was then superseded by 
a table of percentiles for the $h$-index distribution based 
on the number years since one's PhD was (roughly) 
awarded, albeit with several caveats and the use of crude time-interval bins. 
Points (ii) and (iii) were not addressed.  
Abt (2012) tackled issue (i) by dividing the $h$-index by the number of 
decades, or fractions thereof, 
that have elapsed since publication of a researcher’s first paper. 
Here we have endeavoured to correct for all three issues. 
This has been achieved through a rigorous search for research-active
astronomers in Australia, and additionally providing metrics which may 
better reflect an individuals, rather than a team's, performance.  Unlike with career
$h$-indices, there is no need for us to provide plots and tables of the 
$m$-index because our analysis is based on the same fixed 5 and 10 year
interval for all involved.  The results presented here do not however replace
the good work of Pimbblet (2011), but rather build upon it.
As noted previously, 
we have identified an additional 132 PhD-qualified astronomers in Australia. 

The histograms and percentiles presented here provide a valuable snapshot of the 
performance of our nation's astronomers, and also allow individuals to see how
they may be fairing.  Furthermore, it is of interest to see where our 
astronomers are currently distributed across the country (Table~\ref{Tab1} and
Figure~\ref{Fig1}).  With the growth of astronomy, and the associated
publication pressure if one is to remain competitive, comes an increased global
number of citations.  Pearce (2004) showed that, as of November 2003, the top 10\% and
1\% of the world's astronomers based on author-normalised citations over the preceding 5
years (plus 6 months), had 41 and 168 author-normalised citations.   Some 8
years later, the figures for Australian astronomers (albeit allowing an
additional 12 rather than 6 months for citations to accrue, and thus 
a probable $\sim$15\% increase) are 139 and $\sim$400.

\subsection{Total Citations} 

Hirsch (2005, his equation~1) 
found that an author's total number of citations $T$ is proportional to $h^2$, 
with the constant of proportionality ranging from 3 to 5, 
i.e.\ $h \propto \sqrt{T}/\sqrt{4\pm1}$.
Accommodating for the fact that $h=1$ when $T=1$, 
Spruit (2012) has recently argued that in astrophysics the $h$-index 
correlates very tightly with $T$, such that the mean relation is given by 
\begin{equation}
h = 0.5(1+\sqrt{T}) = \frac{1}{\sqrt{4}}(1+\sqrt{T}). 
\end{equation} 
Indeed, his data defining this 
relation showed very little scatter and he concluded that the $h$-index
therefore does not appear to measure anything significantly different to total
citations.
Having additionally collected total citations at the same time as the
$h$-index was collected, we are in a position to explore this claim using
three times more data.  Spruit's self-recognised ``non-random selection
process'' of 113 authors tended to select individuals with $h$-indices
typically greater than 10.  As seen in Table~\ref{Tab2}, over the 10-year
interval from 2001--2010 --- let alone over a 5-year interval --- 
half of our community have not acquired a $h$-index as high as this, 
and it therefore remains to be established how and if the above 
expression applies to much of the community.

\begin{figure}[ht]
\begin{center}
\includegraphics[scale=0.59, angle=-90]{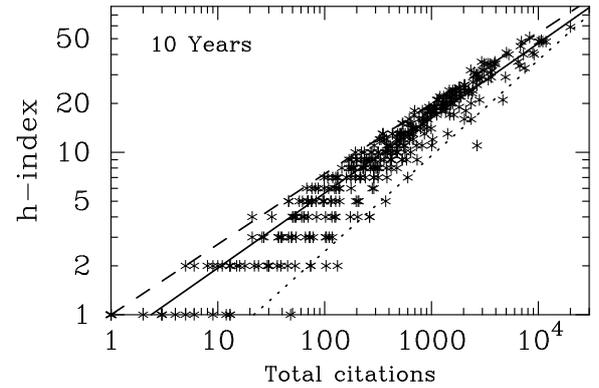}
\caption{
Plot of $h$-indices versus total citation count from publications in
2001–2010, after allowing one year (2011) for citations to accrue.  
The dashed line has been taken from Spruit (2012) 
and is given by $h = (1+T^{1/2})/\sqrt{4}$, where $T$ is the total citation count.
The solid line is given by $h = (0.5+T^{1/2})/\sqrt{5}$, while 
the dotted line is given by $h = (0.6+T^{1/2})/\sqrt{6}$.
} 
\label{Fig4}
\end{center}
\end{figure}

In Figure~\ref{Fig4} we have plotted the $h$-index against the total citation
count, with both sets of values accrued from publications during 2001--2010.
Figure~\ref{Fig4} reveals (i) notably more scatter than Spruit observed, (ii) that 
Spruit's relation is not applicable for $h$ less than $\sim$10, and 
(iii) that his expression defines something of an upper envelope rather 
than the actual mean distribution of our larger and less subjective data set.  Above 
$h\sim10$ we do however observe a dense band similar to that 
found by Spruit (2012), while also 
observing a scatter roughly corresponding to a factor of $\sim$3 in total citation 
count at any given value of $h> \sim10$.  Figure~\ref{Fig4} also reveals that 
our distribution is better described by the mean relation 
\begin{equation}
h = \frac{1}{\sqrt{5}}(0.5+\sqrt{T}), \, \, {\rm for} \, h > \sim 5.
\end{equation}
We find that we are also able to approximate the lower envelope of the distribution by the expression 
\begin{equation}
h = \frac{1}{\sqrt{6}}(0.6+\sqrt{T}), \, \, {\rm for} \, h > \sim 3, 
\end{equation}
which can be considered the counterpart to Spruit's expression which matches the upper
envelope of the distribution. 

As Spruit (2012) noted, the relation between $h$-index and total citation
count, at least above $h\sim$7--10, does indeed reveal that these quantities are
related, which implies that, not surprisingly, one's total citation count is
also likely to be `stretched' in the same way that the $h$-index is due to
membership of large research teams.  Some 20 years ago this was not the
concern that it is today, although it was always recognised that such values
can be disproportionately affected by a single publication of major influence.
The ease of communication via the internet has greatly facilitated the ability
of large numbers of researchers at distant locations to colloborate on and
contribute to large projects (Frogel 2010). Unfortunately, our once favoured metrics are now
something of a reflection of a team's collective performance rather than that
of an individuals contribution and performance.

Spruit (2012) additionally showed that author-normalised citations and
author-normalised $h$-indices are also related by equation~1, and that the
outliers in the $h$--(total citation) diagram are removed when one uses
metrics which have been normalised by the number of contributing authors. 
While we did not collect the data to test it, one may speculate if such a modified 
$h$-index derived from author-normalised citations rather than total citations 
(e.g.\ Batista et al.\ 2006) 
may provide a useful index.  It could be referred to as one's $i$-index,
with the letter $i$ both following alphabetically after the letter $h$ and
better referring to an ``individual'' performance. 
Although, we note that while the validity of the $h$-index may now be
questioned in the new era of large research teams, the $i$-index may also have
its problems, in particular for those who chose to only work in large research
teams.  Therefore, a number of objective, normalised metrics, rather than
solely one, seems likely to persist.  As argued by Panaretos \& Malesios
(2009), these should also be used in combination with other criteria such
as memberships on editorial boards, awards, invitations or peer reviews when
assessing the quality of individual researchers.

%

\section*{Acknowledgments}

The first four authors 
undertook this research / survey as Year 10 work experience students 
at the Swinburne University of Technology. 
The `Astronomers' were identified, and the data was collected, during 
2011 December 6--9 (AP, JT, AZ).  This was checked, and slightly expanded
upon, from 2011 December 12--16 (KHK).  
The text, figures and equations~2 and 3 were provided by AWG, 
with useful comments from 
Duncan A.\ Forbes and Matthew Colless, and helpful links to Scientometrics articles kindly 
provided by Rebecca Parker.  
This work was conducted under the supervision of Duncan A.\ Forbes and AWG. 
AWG is supported by the Australian Research Council through a Future Fellowship
(project FT110100263). The views expressed herein are
those of the authors and are not necessarily those of the Australian Research
Council. 
This research has made use of the SAO/NASA Astrophysics Data System.


\newpage
\pagebreak

\section{Appendix: Tables}

\begin{table}[ht]
\caption{Top ten $h$-indices (2001--2010).\label{Tab3}}
\begin{center}
\begin{tabular}{rlrr}
\hline
\#  &      Author          &  count  & (career) \\ 
\hline
1   & Karl Glazebrook      &   59  & (74) \\  
2   & Gavin Rowell         &   51  & (54) \\    
3   & Bruce Peterson       &   49  & (76) \\    
=4  & Joss Bland-Hawthorn$^*$  &   48  & (60) \\ 
=4  & Matthew Colless      &   48  & (59) \\    
=4  & Warrick Couch        &   48  & (64) \\    
=4  & Simon Driver         &   48  & (53) \\    
=4  & Ken Freeman          &   48  & (80) \\    
5   & Richard Manchester   &   45  & (72) \\    
=6  & Bryan Gaensler       &   41  & (46) \\    
=6  & Carole Jackson       &   41  & (43) \\    
=7  & Russell Cannon       &   40  & (52) \\    
=7  & Geraint Lewis        &   40  & (45) \\    
=7  & Brian Schmidt        &   40  & (54) \\    
=8  & Martin Asplund       &   36  & (42) \\
=8  & David McClelland     &   36  & (38) \\    
=8  & John Norris          &   36  & (61) \\    
=9  & Michael Drinkwater   &   35  & (41) \\    
=9  & Alister Graham       &   35  & (36) \\    
=9  & Gerhardt Meurer      &   35  & (43) \\    
=9  & Lister Staveley-Smith &  35  & (46) \\    
=9  & Chris Tinney         &   35  & (44) \\    
=10 & Duncan Forbes        &   34  & (45) \\    
=10 & Chris Lidman         &   34  & (43) \\    
\hline
\end{tabular}
\end{center}
Twenty-four names make up the top-10 ranking of 
384 Australian astronomers according to 
Hirsch $h$-indices from publications in 2001--2010.
The equals symbol `=' is used here to help 
designate equally-ranked astronomers. 
All authors listed here have been publishing since the start 
of the decade in question, while Cannon and 
Peterson retired but kept publishing (which may have helped or hindered
their productivity). 
The ``career'' $h$-indices were computed on January 31, at the request of the
referee, to provide a helpful frame of reference. 
$^*$The career $h$-index for J.\ Bland-Hawthorn also includes early works published as J.\ Bland.\\
\end{table}


\begin{table}[ht]
\caption{Top 20: Author-normalised citations (2001--2010).\label{Tab4}}
\begin{center}
\begin{tabular}{rlr}
\hline
\# &      Author        &  count \\ 
\hline                       
1  & Martin Asplund       &  1931 \\
2  & Alister Graham       &  1237 \\
3  & Stuart Wyithe	  &  1183 \\
4  & Karl Glazebrook	  &  1097 \\
5  & Joss Bland Hawthorn  &  1067 \\
6  & Ken Freeman 	  &  1013 \\
7  & Richard Manchester	  &  978  \\
8  & Bryan Gaensler	  &  970  \\
9  & Geraint Lewis	  &  964  \\
10  & Andrew Hopkins	  &  936  \\
11 & Duncan Forbes	  &  841  \\
12 & Kenji Bekki	  &  839  \\
13 & Holger Baumgardt	  &  789  \\
14 & Warrick Couch	  &  757  \\
15 & Michael Dopita	  &  736  \\
16 & John Norris	  &  705  \\
17 & Simon Driver	  &  692  \\
18 & Michael Murphy	  &  596  \\
19 & Darren Croton	  &  563  \\
20 & Yuri Levin		  &  530  \\
\hline
\end{tabular}
\end{center}
Author-normalised citations from publications in 2001--2010.  The total
citation count of each article is divided by the number of contributing
authors.\\
\end{table}


\begin{table}[ht]
\caption{Top 20: Lead-author citations (2001--2010).\label{Tab5}}
\begin{center}
\begin{tabular}{rlr}
\hline
\# &      Author        &  count \\ 
\hline
1   &  Martin Asplund     &  3015  \\
2   &  Holger Baumgardt   &  1948  \\
3   &  Stuart Wyithe	  &  1911  \\
4   &  Alister Graham	  &  1712  \\
5   &  Kenji Bekki	  &  1693  \\
6   &  Darren Croton	  &  1567  \\
7   &  Andrew Hopkins	  &  1387  \\
8   &  Michael Murphy	  &  1296  \\
9   &  Bryan Gaensler	  &  1278  \\
10   &  Matthew Colless	  &  1109  \\
11  &  Scott Croom	  &  1011  \\
12  &  Richard Manchester &  994   \\
13  &  Gerhardt Meurer	  &  991   \\
14  &  Chris Blake	  &  803   \\
15  &  George Hobbs	  &  793   \\
16  &  Duncan Galloway	  &  755   \\
17  &  Jarrod Hurley	  &  694   \\
18  &  Daniel Price	  &  669   \\
19  &  Tim Bedding	  &  650   \\
20  &  David Yong	  &  618   \\
\hline
\end{tabular}
\end{center}
\end{table}


\begin{table}[ht]
\caption{Top 20: Author-normalised citations (2006--2010).\label{Tab6}}
\begin{center}
\begin{tabular}{rlr}
\hline
\# &      Author        &  count \\ 
\hline
1   & Martin Asplund      & 566   \\
2   & Darren Croton	  & 468   \\  
3   & Alister Graham	  & 439   \\
4   & Stuart Wyithe	  & 404   \\
5   & Andrew Hopkins	  & 402   \\
6   & Daniel Price	  & 363   \\
7   & Bryan Gaensler	  & 342   \\
8   & Holger Baumgardt    & 339   \\
9   & Richard Manchester  & 308   \\
10   & Kenji Bekki	  & 308   \\
11  & Yuri Levin	  & 298   \\
12  & Geraint Lewis	  & 276   \\
13  & John Norris	  & 250   \\
14  & Duncan Forbes	  & 235   \\
15  & John Lattanzio    & 229   \\
16  & Karl Glazebrook	  & 219   \\
17  & Michael Murphy	  & 218   \\
18  & Quentin Parker	  & 216   \\
19  & Jill Rathborne	  & 206   \\
20  & Ken Freeman 	  & 201   \\
\hline
\end{tabular}
\end{center}
\end{table}


\begin{table}[ht]
\caption{Top 20: Lead-author citations (2006--2010).\label{Tab7}}
\begin{center}
\begin{tabular}{rlr}
\hline
\# &      Author        &  count \\ 
\hline
1   &  Darren Croton	  &  1311  \\  
2   &  Martin Asplund     &  845   \\
3   &  Holger Baumgardt   &  709   \\
4   &  Andrew Hopkins	  &  563   \\
5   &  Alister Graham	  &  502   \\
6   &  Kenji Bekki	  &  482   \\
7   &  Daniel Price	  &  478   \\
8   &  Daniel Zucker	  &  409   \\
9   &  Tamara Davis	  &  363   \\
10   &  James Bolton	  &  334   \\
11  &  Stuart Wyithe	  &  329   \\
12  &  Jill Rathborne	  &  329   \\
13  &  Michael Brown	  &  323   \\
14  &  Bryan Gaensler	  &  322   \\
15  &  Michael Murphy	  &  322   \\
16  &  Amanda Karakas	  &  321   \\
17  &  Simon Driver	  &  318   \\
18  &  Richard Hunstead	  &  315   \\
19  &  Duncan Galloway	  &  311   \\
20  &  David Yong	  &  290   \\
\hline
\end{tabular}
\end{center}
\end{table}


\end{document}